\documentclass[runningheads]{llncs}
\usepackage{tikz}
\usetikzlibrary{arrows,decorations.markings,chains,fit,shapes}
\definecolor{assumptions_asserted}{rgb}{0.7, 0.75, 0.81}
\definecolor{assert}{rgb}{1.0, 0.0, 0.0}
\definecolor{arrow_stream}{rgb}{0, 0, 1.0}
\definecolor{arrow_assumption}{rgb}{0, 0, 1.0}
\usepackage{amsmath}
\usepackage{graphicx}
\usepackage{hyperref}

\usepackage{xcolor}
\usepackage{listings}
\usepackage{lstautogobble}  
\usepackage{xspace} 
\usepackage{todonotes}
\usepackage[]{placeins}
\usepackage{blindtext}
\usepackage{amssymb}
 
\newcommand*\lola{\textsc{Lola}\xspace}
\newcommand*\smt{\textsc{SMT}\xspace}

\makeatletter
\newcommand*{\ie}{i.e.,\@\xspace}
\newcommand*{\eg}{e.g.,\@\xspace}
\makeatother

\definecolor{bluekeywords}{rgb}{0.13, 0.13, 1}
\definecolor{greentypes}{rgb}{0, 0.5, 0}
\definecolor{redstrings}{RGB}{171, 114, 2}
\definecolor{graynumbers}{rgb}{0.5, 0.5, 0.5}
\definecolor{goldcomments}{rgb}{0.6, 0.4, 0.08}
\definecolor{monitorblue}{RGB}{18, 163, 38}

\lstdefinelanguage{Lola}{
  keywords=[0]{input, output, trigger, trigger_once, import, assume, assert, const},
  keywordstyle=[0]\bfseries\color{bluekeywords},
  keywords=[1]{if, then, else, aggregate, defaults, offset},
  keywords=[2]{Int8, Int16, Int32, Int64, UInt8, UInt16, UInt32, UInt64, Bool, Float16, Float32, Float64, @1Hz, @5Hz, @10Hz, @100mHz, @1kHz, a1, a2, a3, a4, a5},
  keywordstyle=[2]\color{greentypes},
    sensitive=false,
    comment=[l]{//},
    morecomment=[s]{/*}{*/},
    morestring=[b]',
    morestring=[b]"
}

\begin{document}
\renewcommand{\thelstlisting}{\arabic{lstlisting}}
\title{Monitoring with Verified Guarantees}
%
%
\author{Johann C.~Dauer\inst{1}\orcidID{0000-0002-8287-2376} \and
Bernd Finkbeiner\inst{2}\orcidID{0000-0002-4280-8441} \and
Sebastian Schirmer\inst{1}\orcidID{0000-0002-4596-2479}}
\authorrunning{Dauer et al.}
%
\institute{German Aerospace Center (DLR), Braunschweig, Germany\\
\email{\{johann.dauer, sebastian.schirmer\}@dlr.de}
\and
Helmholtz Center for Information Security (CISPA), Saarbrücken, Germany
\email{finkbeiner@cispa.saarland}\\
}
\maketitle              
\begin{abstract}
Runtime monitoring is generally considered a light-weight alternative to formal verification. 
In safety-critical systems, however, the monitor itself is a critical component. 
For example, if the monitor is responsible for initiating emergency protocols, as proposed in a recent aviation standard, then the safety of the entire system critically depends on guarantees of the correctness of the monitor.
In this paper, we present a verification extension to the \lola monitoring language that integrates the efficient specification of the monitor with Hoare-style annotations that guarantee the correctness of the monitor specification.  
We add two new operators, assume and assert, which specify assumptions of the monitor and expectations on its output, respectively. 
The validity of the annotations is established by an integrated \smt solver. 
We report on experience in applying the approach to specifications from the avionics domain, where the annotation with assumptions and assertions has lead to the discovery of safety-critical errors in the specifications. 
The errors range from incorrect default values in offset computations to complex algorithmic errors that result in unexpected temporal patterns.


\keywords{Formal methods \and Cyber-physical systems \and Runtime Verification \and Hoare Logic.}
\end{abstract}

\section{Introduction}\label{sec:introduction}
Cyber-physical systems are inherently safety-critical due to their direct interaction with the physical environment -- failures are unacceptable.
A means of protection against failures is the integration of reliable monitoring capabilities.
A \emph{monitor} is a system component that has access to a wide range of system information, \eg sensor readings and control decisions.
When the monitor detects a failure, \ie a violation of the behavior stated in its \emph{specification}, it notifies the system or activates recoveries to prevent failure propagation.

The task of the monitor is critical to the safety of the system, and its correctness is therefore of utmost importance.
Runtime monitoring approaches like \lola~\cite{lola05,VerifiedLola} address this by describing the monitor in a formal specification language, and then generating a monitor implementation that is provably correct and has strong runtime guarantees, for example on memory consumption.
Formal monitoring languages typically feature temporal~\cite{R2U2} and sometimes spatial~\cite{Nenzi} operators that simplify the specification of complex monitoring behaviors. 
However, the specification itself, the central part of runtime monitoring, is still prone to human errors during specification development. 
How can we check that the monitor specification itself is correct?

\begin{figure}
\centering
\resizebox{1.0\textwidth}{!}{%
\begin{tikzpicture}
\draw[very thick] (-0.5,1) rectangle node{External Data} (2,2);
\draw[very thick] (7.5,2.5) rectangle node{Safety Monitor} (10.5,3.5);
\draw[very thick] (3,1) rectangle node{Complex Function} (6,2);

\draw[very thick] (2.9,-0.7) rectangle node{} (7.3,-1.7);
\draw[very thick] (3.2,-0.8) rectangle node{} (7.6,-1.8);
\filldraw[fill=white, very thick] (3.5,-0.9) rectangle node{Recovery Control Function(s)} (7.9,-1.9);

\draw[very thick] (9,0.5) circle (1);
\filldraw[black, very thick] (8,0.5) circle (0.1);	
\filldraw[black, very thick] (9,0.5) circle (0.1);	
\filldraw[black, very thick] (9,1.5) circle (0.1);	
\filldraw[black, very thick] (8.1,0.15) circle (0.1); 
\filldraw[black, very thick] (8.25,-0.15) circle (0.1);	
\filldraw[black, very thick] (8.5,-0.4) circle (0.1);	 
\node[text width=3cm] at (10.1, -0.01) {Switch};

\filldraw[black, very thick] (2.4,1.5) circle (0.05);	
\filldraw[black, very thick] (6.5,1.5) circle (0.05);	
\filldraw[black, very thick] (2.4,-1.1) circle (0.05);	
\filldraw[black, very thick] (2.4,-1.25) circle (0.05);	

\draw [>=stealth, very thick] (2,1.5) -- (2.4,1.5);	
\draw [>=stealth, very thick] (2.4,.5) -- (2.4,3.15); 	
\draw [->,>=stealth, very thick] (2.4,3.15) -- (7.5,3.15);	
\draw [->,>=stealth, very thick] (9,2.5) -- (9,1.6);	
\draw [->,>=stealth, very thick] (2.4,1.5) -- (3,1.5);	
\draw [>=stealth, very thick] (9,1.5) -- (9,0.5);	
\draw [->,>=stealth, very thick] (9,0.5) -- (8.15,0.5);	
\draw [->,>=stealth, very thick] (10,0.5) -- (11,0.5);	
\draw [->,>=stealth, very thick] (2.4,1.5) -- (3,1.5);	

\draw [>=stealth, very thick] (2.4,.5) -- (2.4,-1.4); 
\draw [->,>=stealth, very thick] (2.4,-1.1) -- (2.9,-1.1); 
\draw [->,>=stealth, very thick] (2.4,-1.25) -- (3.2,-1.25); 
\draw [->,>=stealth, very thick] (2.4,-1.4) -- (3.5,-1.4); 

\draw [>=stealth, very thick] (6,1.5) -- (6.5,1.5);	
\draw [>=stealth, very thick] (6.5,1.5) -- (6.5,0.5); 
\draw [->,>=stealth, very thick] (6.5,0.5) -- (7.9,0.5);	
\draw [>=stealth, very thick] (6.5,1.5) -- (6.5,2.85);	
\draw [->,>=stealth, very thick] (6.5,2.85) -- (7.5,2.85);	

\draw [>=stealth, very thick] (5.2,-0.7) -- (5.2,0.15);	
\draw [->,>=stealth, very thick] (5.2,0.15) -- (8,0.15); 
\draw [>=stealth, very thick] (5.4,-0.8) -- (5.4,-0.15);	
\draw [->,>=stealth, very thick] (5.4,-0.15) -- (8.15,-0.15); 
\draw [>=stealth, very thick] (5.6,-0.9) -- (5.6,-0.4);	
\draw [->,>=stealth, very thick] (5.6,-0.4) -- (8.4,-0.4); 
\end{tikzpicture}
}%
\caption{Run-Time Assurance architecture proposed by ASTM F3269 to safely bound a complex function using a safety monitor.}
\label{fig:astm}
\centering
\end{figure}
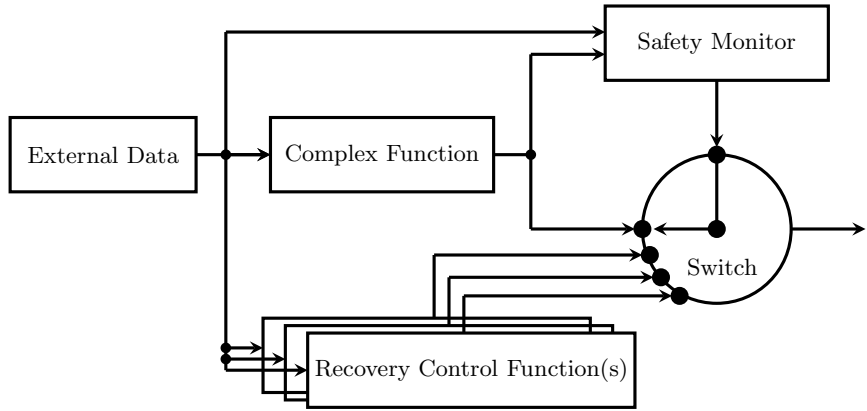

In this paper, we introduce a verification feature to the \lola framework.
Specifically, we extend the specification language with \emph{assumptions} and \emph{assertions}. 
The framework verifies that the assertions are guaranteed to hold if the input to the monitor satisfies the assumptions.
The prime application area of \lola is unmanned aviation.
\lola is increasingly used for the development and operation monitoring of unmanned aircraft; for example, the \lola monitoring framework has been integrated into the DLR unmanned aircraft superARTIS\footnote{https://www.dlr.de/content/en/research-facilities/superartis-en.html}~\cite{fpgartlola}.
The verification extension presented in this paper is motivated by this work.
In practice, system engineers report that support for specification development is necessary, \eg sanity checks and proves of correctness.
Additionally, recent developments in unmanned aviation regulations and standards indicate a similar necessity.
One such development is the upcoming industry standard ASTM F3269 (Standard Practice for Methods to Safely Bound Flight Behavior of Unmanned Aircraft Systems Containing Complex Functions).
ASTM F3269 introduces a certification strategy based on a 
Run-Time Assurance (RTA) architecture that bounds the behavior of a complex function by a safety monitor~\cite{astm}, similar to the well-known Simplex architecture~\cite{sha}. 
This complex function could be a Deep Neural Network as proposed in~\cite{codann}. 
A simplified version of the architecture\footnote{In its original version the data is separated into assured and unassured data and data preparation components are added.} of ASTM F3269 is shown in Figure~\ref{fig:astm}.

At the core of the architecture is a safety monitor that takes the inputs and outputs of the complex function, and decides whether the complex function behaves as expected.
If not, the monitor switches the control from the complex function to a matching recovery function.
For instance, the flight of an unmanned aircraft could be separated into different phases: \eg take-off, cruise flight, and landing.
For each of these phases, a dedicated recovery could be defined, \eg braking during take-off, the activation of a parachute during cruise flight, or a go-around maneuver during landing.
Further, it is crucial that recoveries are only activated under certain conditions and that only one recovery is activated at a time.
For instance, a parachute activation during a landing approach is considered safety-critical.
The verification extension of \lola introduced in this paper can be used to guarantee statically that such decisions are avoided within the monitor specification.
Consider the simplified \lola specification
\begin{lstlisting}[]
input event_a, event_b, value: Bool, Bool, Float32
assume <a1> !(event_a and event_b)
output braking$~$ : Bool := ...computation...
output parachute : Bool := ...computation...
output go_around : Bool := ...computation...
assert <a1> !(braking and parachute) 
\end{lstlisting}
that declares an assumption on the system input \texttt{event}s and asserts that \texttt{braking} and \texttt{parachute} never evaluates to \emph{true} simultaneously.

In the following, we first give a brief introduction to the stream-based specification language \lola, then present the verification approach, and, finally, give details on the tool implementation and our tool experience with specifications that were written based on interviews with aviation experts.
Our results show that standard \lola specifications are indeed prone to error, and that these errors can be caught with the formal verification introduced by our extension. 

\paragraph{\textbf{Related Work}}\label{sec:related work}~\\
Most work on the verification of monitors focuses on the correct transformation into a general programming language.
For example, Copilot~\cite{copilot} specifications can be compiled into C code with constant time and memory requirements.
Similarly, there is a translation validation toolkit for \lola monitors implemented in Rust~\cite{VerifiedLola}, which is based on the Viper verification tool. Translation validation of this type is orthogonal to the verification approach of this paper. 
Instead of verifying the correctness of a transformation, our focus is to verify the specification itself.
Both activities complement each other and facilitate safer future cyber-physical systems.

Our verification approach is based on classic ideas of inductive program verification~\cite{hoare69,Floyd1993}, and is closely related to the techniques used in static program verifiers like \textsc{KeY}~\cite{KeYBook2007}, VeriFast~\cite{verifast}, and Dafny~\cite{dafny}.
In a verification approach like Dafny, we are interested in functional properties of procedures,
specified as post-conditions that relate the values upon the termination of the procedure with those at the time of entry to the procedure, \eg \emph{ensure y = old(y)}.
By contrast, a stream-based language like \lola allows arbitrary access to past and future stream values.
This makes it necessary to \emph{unfold} the \lola specification in order to properly relate the assumptions and assertions in time.

Most closely related to stream-based monitoring languages are synchronous programming languages like \textsc{LUSTRE}~\cite{lustre}, \textsc{ESTEREL}~\cite{esterel}, and \textsc{SIGNAL}~\cite{signal}.
For these languages, the compiler is typically used for verification -- a program representing the negation of desired properties is compiled with the target program and a check for emptiness decides whether the properties are satisfied.
Furthermore, a translation from past linear-time temporal logic to \textsc{ESTEREL} was proposed to simplify the specification of more complex temporal properties~\cite{esterelPLTL}.
Other verification techniques also exist like \smt-based \emph{k-}Induction for \textsc{LUSTRE}~\cite{smtbasedlustre} or a term rewriting system on synced effects \cite{trsesterel}.
A key difference in our approach is that we do not rely on compilation. 
Our verification works on the level of an intermediate representation.
Furthermore, synchronous programming languages are limited to past references, while the stream unfolding for the inductive correctness proof of the \lola specification includes both past and future temporal operators.
Similar to \emph{k-}Induction, our approach is sound but not complete.

\section{Runtime Monitoring with \lola}\label{sec:preliminaries}
%

We now give an overview of the monitoring specification language \lola.
The verification extension is presented in the next section.

\lola is a stream-based language that describes the translation from input streams to output streams:
{
\setlength\abovedisplayshortskip{0pt}
\setlength\belowdisplayshortskip{0pt}
\setlength\abovedisplayskip{5pt}
\setlength\belowdisplayskip{-5pt}
\begin{align*} 
\text{\textbf{input}}~ t_1  &: T_1\\
\vdots &\\
\text{\textbf{input}}~ t_m  &: T_m\\
\text{\textbf{output}}~ s_1 &: T_{m+1} := e_1(t_1,\dots,t_m,s_1, \dots, s_n)\\
\vdots &\\
\text{\textbf{output}}~ s_n &: T_{m+n} := e_n(t_1,\dots,t_m,s_1, \dots, s_n)\\
\text{\textbf{trigger}}~ \varphi & ~ \mathit{message}\\
\end{align*}
}
where input streams carry synchronous arriving data from the system under scrutiny, output streams represent calculations, and triggers generate notification $\mathit{message}$s at instants where their condition $\varphi$ becomes $\mathit{true}$.
Input streams $t_1,\dots, t_m$ and output streams $s_1,\dots, s_n$ are called \emph{independent} and \emph{dependent variables}, respectively.
Each variable is typed: independent variables $t_i$ are typed $T_i$ and dependent variables $s_i$ are typed $T_{m+i}$.
Dependent variables are computed based on \emph{stream expressions} $e_1, \dots, e_n$ over dependent and independent stream variables.
A stream expression is one of the following:
\begin{itemize}
	\item an atomic stream expression $c$ of type $T$ if $c$ is a constant of type $T$;
	\item an atomic stream expression $s$ of type $T$ if $s$ is a stream variable of type $T$;
	\item a stream expression $ite(b, e_1, e_2)$ of type $T$ if $b$ is a Boolean stream expression and $e_1, e_2$ are stream expressions of type $T$. Note that $ite$ abbreviates the control construct \emph{if-then-else};
	\item a stream expression $f(e_1, \dots, e_k)$ of type $T$ if $f: T_1 \times \dots \times T_k \mapsto T$ is a $k$-ary operator and $e_1, \dots, e_k$ are stream expressions of type $T_1, \dots, T_k$;
	\item a stream expression $\mathit{o.offset(by: i).defaults(to: d)}$ of type $T$ if $o$ is a stream variable of type $T$, $i$ is an Integer, and $d$ is of type $T$.
\end{itemize}

\noindent For example, consider the \lola specification
\begin{lstlisting}[]
input   altitude: Float32 // in m
output  altitude_bound := altitude > 200.0
trigger altitude_bound "Warning: Decrease altitude!"
\end{lstlisting}
that notifies the system if the current \texttt{altitude }is above its operating limits, \ie \texttt{200.0} meters.
Note that stream types are inferred, \ie \texttt{altitude\_bound} is of type \texttt{Bool}. 

\lola uses temporal operators that allow output streams to access its and others previous and future stream values.
The stream 
\begin{lstlisting}[]
output alt_count := if altitude $\le$ 200.0 then 0
 		            else alt_count.offset(by: -1).defaults(to: 0) + 1
\end{lstlisting}
represents a count of consecutive altitude violations by accessing its own previous value, \ie \texttt{offset(by: x)} where a negative and positive integer \texttt{x} represents past and future stream accesses, respectively.
Since temporal accesses are not always guaranteed to exist, the default operator defines values which are used instead, \ie \texttt{defaults(to: d)} where \texttt{d} has to be of the same type as the used stream.
Here, at the first position of \texttt{alt\_count} the default value zero is taken.
As abbreviations for the temporal operators, \texttt{alt\_count[x, d]} is used.
Further, \texttt{s[x..y, d, $\circ$]} for \texttt{x $<$ y} abbreviates \texttt{s[x,d] $\circ$ s[x+1,d] $\circ~\dots~\circ$ s[y,d]} where $\circ$ is a binary operator.
Using \texttt{alt\_count > 10} as a trigger condition is preferable if only persistent violations should be reported.

In general, \lola is a specification language that allows to specify complex temporal properties in a precise, concise, and less error-prone way.
The focus is on \emph{what} properties should be monitored instead of \emph{how} a monitor should be executed.
Therefore, the \lola monitor synthesis automatically infers and optimizes implementation details like evaluation order and memory management.
The evaluation order~\cite{VerifiedLola} of \lola streams is automatically derived by analysis of the \emph{dependency graph}~\cite{lola05} of the specification.
This allows to ignore the order when taking advantage of the modular structure of \lola output streams, \eg:
\begin{lstlisting}[]
output alt_avg := alt_count / (position+1)
output alt_count := if altitude $\le$ 200.0 then 0
 		            else alt_count.offset(by: -1).defaults(to: 0) + 1
output position := position.offset(by: -1).defaults(to: 0)
\end{lstlisting}
where \texttt{position} and \texttt{alt\_count} are used before their definition.
Further, the dependency graph allows to detect invalid cyclic stream dependencies, \eg 

\noindent \texttt{output a := a.offset(by: 0).defaults(to: 0)}.

\section{Assumptions and Assertions}\label{sec:hoarelola}
In this section, we present the verification extension for the \lola specification language.
The extension allows the developer to annotate the \lola specification with \emph{assumptions} and \emph{assertions} in order to verify the desired guarantees on the computed streams.
As an example, consider the simplified specification in Listing \ref{lst:recovery}, which is structured into stream computations in Lines 1 to 23, and assumptions and assertions from Line 26 onwards.

\vspace{5mm}
\lstset
{ 
    numbers=right,
    stepnumber=1,
}
\begin{lstlisting}[caption={A simplified Run-Time Assurance \lola specification with three recovery functions for three different flight phases. Assumptions and assertions are used to show that only one recovery function is activated at once.}, label={lst:recovery}]
input alt :  Float32 // Height above ground
input x, y : Float32, Float32 // Position in local coordinate system
input speed : Float32  // Velocity of aircraft
input landing : Bool // Indicates landing mode
input lg_status : (Float32,Float32,FLoat32) // Status of landing gear

// Complex computations
output dst_on_runway $~~~~~$: Float32 := $\sqrt{x^2 + y^2}$
output geofence_violation : Bool $~~$ := ...
output landing_gear_ready : Bool $~~~$:= ...

// Take-off contingency 
output decelerate := alt $<$ 1.0 $\wedge$ speed $<$ 10.0 $\wedge$ dst_on_runway > 20.0 
// In-flight contingency
output parachute := geofence_violation $\wedge$ alt $>$ 100.0 
// Landing contingency
output gain_alt := landing $\wedge$ alt $\ge$ 10.0 $\wedge$ (speed $>$ 10.0 $\vee$
                     !landing_gear_ready[-4..0, true, $\wedge$])
            
// Notifications to the system
trigger decelerate "RECOVERY: Stop take-off by decelerating aircraft."
trigger parachute $~$"RECOVERY: Activate parachute."
trigger gain_alt $~~$"RECOVERY: Gain altitude for next landing attempt."

// By concept of operations: landing is always within geofence.
assume <a1> $\neg$(landing $\wedge$ geofence_violation)
assume <a1> abs(speed) <= 80.0 // Given by data protocol

// Only one contingency is activated at once.
assert <a1> $\neg$( (decelerate $\wedge$ parachute)  $\vee$  (decelerate $\wedge$ gain_alt)
               $\vee$  (parachute  $\wedge$ gain_alt) )  
// Parachute SHALL ONLY be activated 100 m above ground.
assert <a2> parachute $\rightarrow$ alt > 100.0  
\end{lstlisting}

\newpage

The computation part specifies a safety monitor within a RTA architecture that triggers recovery functions for three different flight phases.
First, the take-off recovery function is triggered (Line 21) when the targeted take-off speed was not achieved on a runway up to a predefined point (Line 13).
The distance between the current position and the end of the runway with local coordinates $(0,0)$ is computed in Line 8.
Second, in-flight a parachute is activated (Line 22) when virtual barriers for the aircraft, \ie a geofence, are exceeded (Line 15). 
For more details on a \lola geofence specification (Line 9), we refer to~\cite{geofences}.
Last, during landing, up to a point of no return (\texttt{alt < 10.0}), a new landing attempt is initiated (Line 23) if the aircraft's speed is too fast or its landing gear is not yet ready.
To be more robust, the current and the previous value of the \texttt{landing\_gear\_ready} is taken into account (Lines 17-18).

With the verification extension, the specification assures that recoveries are not activated simultaneously (Lines 30-31), \ie for instance there is no possibility that a parachute is activated during a landing approach.
The first two conjunctions in Line 30 evaluate to $\mathit{false}$ because relevant outputs use a disjoint altitude condition.   
The last conjunction requires an assumption.
In fact, here, two assumptions are linked by the identifier $a1$ to the assertion. 
The assumptions specify: the known bound of received speed data (Line 27) as well as operational information (Line 26), \eg given by the concept of operation a nominal landing is only foreseen within the predefined operational airspace.
Further, a second assertion is stated in Line 33 that guarantees that \emph{the parachute should only be activated when the aircraft is 100 meters above ground}.
In this case, the property can be shown assumption-free.
Assertions help engineers to show that certain properties are $\mathit{true}$.
The given assertions indicate how specification debugging and management can benefit from the extension -- it avoids digging into potentially complex stream computations.

The extension and its verification approach are presented in the following.
In general, the verification extension is used if a \lola specification is annotated in the following way:
{
\setlength\abovedisplayshortskip{0pt}
\setlength\belowdisplayshortskip{0pt}
\setlength\abovedisplayskip{5pt}
\setlength\belowdisplayskip{-5pt}
\begin{align*}
\text{\textbf{assume}}~ \langle \alpha_1 \rangle &\quad \theta_1\\
\vdots &\\
\text{\textbf{assume}}~ \langle \alpha_m \rangle &\quad \theta_m\\
\text{\textbf{assert}}~ \langle \alpha_{m+1} \rangle &\quad \psi_1\\
\vdots &\\
\text{\textbf{assert}}~ \langle \alpha_{m+n} \rangle &\quad \psi_n\\
\end{align*}
}
where $\alpha_1,\dots,\alpha_{m+n} \in \Gamma$ are identifiers for $\theta_1,\dots,\theta_m,\psi_1,\dots,\psi_n$, which are Boolean stream expressions with possibly temporal operators.
For convenience, we define functions which return all $\theta$ and $\psi$ that are linked to a given $\alpha$ identifier:\\
 $assume(\alpha) = \{ \theta_j ~|~ \forall \alpha_j \in \Gamma, \alpha = \alpha_j\}$ and $assert(\alpha) = \{ \psi_j ~|~ \forall \alpha_j \in \Gamma, \alpha = \alpha_j\}$.
The set of assertion $\psi_1,\dots,\psi_n$ is \emph{correct} for all input streams iff whenever an assumption is satisfied, its corresponding assertion is satisfied as well.

The verification of assertions relies on the encoding of the \lola execution in Satisfiability Modulo Theory (\smt).
We define the $smt$ function that encodes a stream expression next.
It can be used to encode independent and dependent variables as well as expressions of assumptions and assertions.

\begin{definition}[\smt-Encoding of Stream Expressions]\label{def-encoding}\\
Let $\Phi$ be a \lola specification over independent stream variables $t_1,\dots,t_m$ and dependent stream variables $s_1,\dots,s_n$. 
Further, let the natural number $N+1$ be the length of the input streams, $c$ be an \smt constant symbol, and $\tau_1^0,\dots,\tau_1^N, \dots,  $ $\tau_m^0, \dots, \tau_m^N,~ \sigma_1^0, $ $\dots, \sigma_1^N, \dots, \sigma_n^0, \dots, \sigma_n^N$ be \smt variables. 
Then, the function $smt$ recursively encodes a stream expression $e$ at position j with $0 \leq j \leq N$ in the following way:
\begin{itemize}
\item Base cases:
\begin{itemize}
	\item $smt(c)(j) = \mathtt{c}$
	\item $smt(t_i)(j) = \tau_i^j$
	\item $smt(s_i)(j) = \sigma_i^j$
\end{itemize}
\item Recursive cases:
\begin{itemize}
	\item $smt(f(e_1, \dots, e_n))(j) = \mathtt{f}(smt(e_1)(j),\dots,~smt(e_n)(j))$
	\item $smt(ite(e_b, e_1, e_2))(j) = \mathtt{ite}(smt(e_b)(j),~smt(e_1)(j),~smt(e_2)(j))$
	\item $
  smt(e[k,c])(j) =
  \begin{cases}
    smt(e)(j+k) & \text{if}~ 0 \leq j +k \leq N,\\
    c & \text{otherwise}
  \end{cases}
$
\end{itemize}
\end{itemize}
where $\mathtt{ite}$ is an \smt encoding of \emph{if-then-else}; $\mathtt{f}$ is an interpreted function if $f$ is from a theory supported by the \smt solver and an uninterpreted function otherwise.
\end{definition}

Next, Proposition~\ref{the-hoare} shows how the correctness of asserted stream properties can be proven for finite input streams. 
If the set of assertions is correct, asserted stream properties are guaranteed to be valid in each step of the monitor execution.
In practice, such specifications are preferable.
In the following, let $\Phi$ be a \lola specification with verification annotations.
Further, we refer to the set of input streams and computed output streams as stream execution.

\begin{proposition}[Assertion Verification of a Finite Stream Execution]\label{the-hoare}~\\
The set of assertions is correct for a finite stream execution with length $N+1$ under given assumptions, if the following formula is valid:

\vspace{3mm}
$\bigwedge\limits_{i:~0 \le i \leq N} ~~\Bigl(~~\bigwedge\limits_{\alpha \in \Gamma}~~\Bigl(\\
\bigwedge\limits_{\theta~\in~\mathit{assume}(\alpha)} smt(\theta)(i) ~ \wedge \bigwedge\limits_{s_k \in \Phi} \sigma_k^i = smt(e_k)(i) \rightarrow \bigwedge\limits_{\psi~\in~\mathit{assert}(\alpha)} smt(\psi)(i)~~\Bigr)\Bigr)$
\vspace{3mm}
\end{proposition}

The formula in Proposition~\ref{the-hoare} unfolds the complete stream execution and informally expresses that an assertion must hold in each stream position whenever its corresponding assumption and implementation are satisfied.
 
To avoid the complete unfolding and allow arbitrary stream lengths, an inductive argument is given in Proposition~\ref{def-efficient} that defines proof obligations for an annotated \lola specification.
Next, we present a template for the stream unfolding that helps to define the proof obligation at the \emph{Begin}ning (Definition~\ref{def_begin}), during \emph{Run} (Definition~\ref{def_run}), and at the \emph{End} (Definition~\ref{def_end}) of a stream execution.

\begin{definition}[Template Stream Unfolding]\label{template}~\\
We define the template formula $\phi_{t}$ that states proof obligations as:
\vspace{3mm}

\noindent$\bigwedge\limits_{\alpha \in \Gamma}\Biggl($
$\bigwedge\limits_{i:~\mathit{c\_asm}}\Bigl(~\bigwedge\limits_{\theta~\in~\mathit{assume}(\alpha)} smt(\theta)(i)\Bigr) ~ \wedge ~ $
$\bigwedge\limits_{i:~\mathit{c\_asserted}}\Bigl(~\bigwedge\limits_{\psi~\in~\mathit{assert}(\alpha)} smt(\psi)(i)\Bigr)$\vspace{3mm}\\
$\wedge \bigwedge\limits_{i:~\mathit{c\_streams}}\Bigl(~\bigwedge\limits_{0 < k \leq n} \sigma_k = smt(e_k)(i) \Bigr) ~ \rightarrow ~ $
$\bigwedge\limits_{i:~\mathit{c\_assert}}\Bigl(~\bigwedge\limits_{\psi~\in~\mathit{assert}(\alpha)} smt(\psi)(i)\Bigr)\Biggr)$
\vspace{3mm}

\noindent where $\mathit{c\_asm}$, $\mathit{c\_asserted}$, $\mathit{c\_streams}$, and $\mathit{c\_assert}$ are template parameters for the unfolding of assumptions, previously proven assertions, output streams, and assertion, respectively.
\end{definition}

The template formula in Definition~\ref{template} uses template parameters for the stream unfolding.
For instance, the parameter assignment $\mathit{c\_asm}:=0\leq i < 10$ adds assumptions at the first ten positions of the stream execution. 
Further, the parameter $\mathit{c\_asserted}$ allows to incorporate the induction hypothesis.

In the following, we will use the \lola specification
\begin{lstlisting}[numbers=none]
assume<a1> reset[-1, f] $\vee$ reset[1, f]
input reset : Bool
output o1 := if reset then  0 else o1[-1, 0] + 1 
output o2 := o1[-1, 0] + o1 + o1[1, 0]
assert<a1> 0 $\le$ o2 and o2 $\le$ 3
\end{lstlisting}
as a running example for the template stream unfolding.
Here, the input $\mathit{reset}$ represent a reset command for the output stream $\mathit{o1}$ that counts how long no $\mathit{reset}$ occurred.
Output $\mathit{o1}$ is used by output $\mathit{o2}$ which aggregates over the previous, the current, and the next outcome of $\mathit{o1}$.
As assertion, we show that $\mathit{o2}$ is always positive and never larger than three given the assumption that in each execution step either the previous or the next $\mathit{reset}$ is $\mathit{true}$.
The assumption ensures that at most two consecutive $\mathit{resets}$ are $\mathit{false}$.
Given the $\mathit{reset}$ sequence of input values $\langle \mathit{true}; \mathit{false}; \mathit{false}\rangle$ that satisfies the assumption, the resulting $\mathit{o1}$ stream evaluates to $\langle 0; 1; 2 \rangle$.
Here, at the second position of the sequence, $\mathit{o2}$ evaluates to three.
To show that the assertion also holds at the first and the last position of the sequence, out-of-bounds values must be considered.

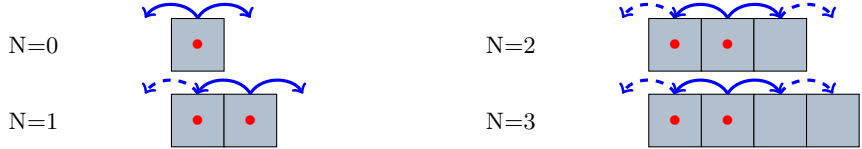
\begin{figure}[b]
\centering
\begin{minipage}{.48\textwidth}
\begin{tikzpicture}
\edef\sizetape{0.7cm}
\tikzstyle{entry}=[draw,minimum size=\sizetape]
\begin{scope}[start chain=1 going right,node distance=-0.15mm]
    \node [on chain=1] {N=0};
    \node [on chain=1, entry, draw=none] (empty0){};
    \node [on chain=1, entry, draw=none] (empty1){};
    \node [on chain=1, entry, fill=assumptions_asserted] (node0){\textcolor{assert}{$\bullet$}};
    \node [on chain=1, entry, draw=none] (node1){};
    \node [on chain=1, entry, draw=none] (node2){};
    \node [on chain=1, entry, draw=none] (node3){};
\end{scope}
\draw[very thick,->, bend right=60, color=arrow_stream] (node0.north) to (empty1.north);
\draw[very thick,->, bend left=60, color=arrow_stream] (node0.north) to (node1.north);
\end{tikzpicture}

\begin{tikzpicture}
\edef\sizetape{0.7cm}
\tikzstyle{entry}=[draw,minimum size=\sizetape]
\begin{scope}[start chain=1 going right,node distance=-0.15mm]
    \node [on chain=1] {N=1};
    \node [on chain=1, entry, draw=none] (empty0){};
    \node [on chain=1, entry, draw=none] (empty1){};
    \node [on chain=1, entry, fill=assumptions_asserted] (node0){\textcolor{assert}{$\bullet$}};
    \node [on chain=1, entry, fill=assumptions_asserted] (node1){\textcolor{assert}{$\bullet$}};
    \node [on chain=1, entry, draw=none] (node2){};
    \node [on chain=1, entry, draw=none] (node3){};
\end{scope}
\draw[very thick,->, bend right=60, color=arrow_assumption, dashed] (node0.north) to (empty1.north);
\draw[very thick,->, bend right=60, color=arrow_stream] (node1.north) to (node0.north);
\draw[very thick,->, bend left=60, color=arrow_stream] (node1.north) to (node2.north);
\end{tikzpicture}
\end{minipage}
\hfill
\begin{minipage}{.48\textwidth}

\begin{tikzpicture}
\edef\sizetape{0.7cm}
\tikzstyle{entry}=[draw,minimum size=\sizetape]
\begin{scope}[start chain=1 going right,node distance=-0.15mm]
    \node [on chain=1] {N=2};
    \node [on chain=1, entry, draw=none] (empty0){};
    \node [on chain=1, entry, draw=none] (empty1){};
    \node [on chain=1, entry, fill=assumptions_asserted] (node0){\textcolor{assert}{$\bullet$}};
    \node [on chain=1, entry, fill=assumptions_asserted] (node1){\textcolor{assert}{$\bullet$}};
    \node [on chain=1, entry, fill=assumptions_asserted] (node2){};
    \node [on chain=1, entry, draw=none] (node3){};
\end{scope}
\draw[very thick,->, bend right=60, color=arrow_assumption, dashed] (node0.north) to (empty1.north);
\draw[very thick,->, bend right=60, color=arrow_stream] (node1.north) to (node0.north);
\draw[very thick,->, bend left=60, color=arrow_stream] (node1.north) to (node2.north);
\draw[very thick,->, bend left=60, color=arrow_assumption, dashed] (node2.north) to (node3.north);
\end{tikzpicture}

\begin{tikzpicture}
\edef\sizetape{0.7cm}
\tikzstyle{entry}=[draw,minimum size=\sizetape]
\begin{scope}[start chain=1 going right,node distance=-0.15mm]
    \node [on chain=1] {N=3};
    \node [on chain=1, entry, draw=none] (empty0){};
    \node [on chain=1, entry, draw=none] (empty1){};
    \node [on chain=1, entry, fill=assumptions_asserted] (node0){\textcolor{assert}{$\bullet$}};
    \node [on chain=1, entry, fill=assumptions_asserted] (node1){\textcolor{assert}{$\bullet$}};
    \node [on chain=1, entry, fill=assumptions_asserted] (node2){};
    \node [on chain=1, entry, fill=assumptions_asserted] (node3){};
\end{scope}
\draw[very thick,->, bend right=60, color=arrow_assumption, dashed] (node0.north) to (empty1.north);
\draw[very thick,->, bend right=60, color=arrow_stream] (node1.north) to (node0.north);
\draw[very thick,->, bend left=60, color=arrow_stream] (node1.north) to (node2.north);
\draw[very thick,->, bend left=60, color=arrow_assumption, dashed] (node2.north) to (node3.north);
\end{tikzpicture}
\end{minipage}
\caption{Four stream executions of different length $N+1$ with the respective template unfolding are depicted.
The stream executions consider all cases with past out-of-bound accesses. 
A gray-colored box indicates that an assumption has been unfolded at this position, while a red dotted box indicates that an assertion has been unfolded at this position. 
Solid and dashed arrows indicate accesses by streams and annotations, respectively.}
\label{fig:starttrace}
\end{figure}

We show how the template $\phi_{t}$ can be used at the beginning of a stream execution.
Here, default values due to past stream accesses beyond the beginning of a stream need to be captured by the obligation to guarantee that the assertions hold in these cases.
The combination of past out-of-bounds and future out-of-bounds default values must also be covered by the obligations in case the stream is stopped early.
These scenarios are depicted for the running example in Figure~\ref{fig:starttrace}.
The figure shows four finite stream executions with different lengths.
All stream positions are colored gray, while only some positions contain a single red dot.
These features indicate the unfolding of stream variables and annotations using the template $\phi_{t}$. 
A gray-colored position means that the assumptions have been unfolded and a dotted position means the assertion has been unfolded.
Further, arrows indicate temporal stream accesses where solid lines correspond to accesses by outputs and dashed lines correspond to accesses by annotations, \ie assumptions and assertions.
For each stream execution, only the arrows for a single position are depicted -- the arrows for other positions have been omitted for the sake of clarity.
For example, for $N=0$, the accesses of output $o2$ are both out-of-bounds, \ie the default value zero is used.
While for $N=3$, the accesses at the second position are shown where only the past access of the assumption leads to an out-of-bounds access.     
The figure depicts all necessary stream execution that cover all combinations of past out-of-bounds accesses, \ie with and without future bound violations.
The described unfoldings of Figure~\ref{fig:starttrace} are formalized as proof obligations in Definition~\ref{def_begin}.

\begin{definition}[Proof Obligations for Past Out-of-bounds Accesses]\label{def_begin}~\\
Let $w_p = \sup( \{0\} \cup \{~\left|k\right|~|~e[k,c] \in \Phi\ $ where $k < 0 \} )$ be the most negative offset and $w_f = \sup( \{0\} \cup \{~k~|~e[k,c] \in \Phi\ $ where $k > 0 \} )$ be the greatest positive offset. 
The proof obligations $\phi_{\mathit{Begin}}$ for past out-of-bounds accesses are defined as the conjunction of template formulas:
\vspace{3mm}

$\bigwedge\limits_{N: ~ 0 \leq N < \max(1,~ 2 \cdot (w_p + w_f))} ~ \phi_{t}(\mathit{c\_asm}, ~\mathit{c\_asserted}, ~\mathit{c\_streams}, ~\mathit{c\_assert})$
\vspace{3mm}

\noindent with template parameters:\\
	\begin{tabular}{ll}
		$~\bullet~ \mathit{c\_asm}$ & $:= 0 \leq i \le N$,\\
		$~\bullet~ \mathit{c\_asserted}$ & $:= \mathit{false}$,\\
		$~\bullet~ \mathit{c\_streams}$ & $:= 0 \leq i \le N$,\\
		$~\bullet~ \mathit{c\_assert}$ & $:= 0 \leq i < \max(1,~\min(N+1,~ 2 \cdot w_p)).$
	\end{tabular}
\end{definition}

\newpage
Next, the case where no out-of-bounds access occurs is considered.
Hence, the obligations capture the nominal case where no default value is used.
Since we have shown that past out-of-bounds accesses are valid we can use these proven assertions as assumptions.
Figure~\ref{fig:runtrace} depicts a stream execution with a single dotted position, \ie the position where the assertion must be proven.
As can be seen, all accesses from this position are within bounds.
Further, note that the accesses of the first and the last unfolded assumption, \ie the first and the last gray-colored position, are also within bounds.   
The described unfolding is formalized as proof obligations in Definition~\ref{def_run}.

\begin{definition}[Proof Obligations for No Out-of-bounds Accesses]\label{def_run}~\\
The proof obligations $\phi_{\mathit{Run}}$ without out-of-bounds accesses are defined as\\ $\phi_{t}(\mathit{c\_asm}, ~\mathit{c\_asserted}, ~\mathit{c\_streams}, ~\mathit{c\_assert})$ with template parameters:\\
\begin{tabular}{ll}
		 $~\bullet~ \mathit{c\_asm}$ & $:= w_p \leq i \leq N - w_f$,\\
		 $~\bullet~ \mathit{c\_asserted}$ & $:= 2 \cdot w_p \leq i \leq N - 2\cdot w_f \wedge i \neq 3\cdot w_p$,\\
		 $~\bullet~ \mathit{c\_streams}$ & $:= 2 \cdot w_p \leq i \leq N - 2\cdot w_f$,\\
		 $~\bullet~ \mathit{c\_assert}$ & $:= i = 3\cdot w_p$,
	\end{tabular}
	
\noindent where $N = 3  \cdot (w_p + w_f)$.
\end{definition}

\begin{figure}
\centering
\begin{tikzpicture}
\edef\sizetape{0.7cm}
\tikzstyle{entry}=[draw,minimum size=\sizetape]
\begin{scope}[start chain=1 going right,node distance=-0.15mm]
    \node [on chain=1] {N=4};
    \node [on chain=1, entry, draw=none] (empty0){};
    \node [on chain=1, entry, draw=none] (empty1){};
    \node [on chain=1, entry] (node0){};
    \node [on chain=1, entry, fill=assumptions_asserted] (node1){};
    \node [on chain=1, entry, fill=assumptions_asserted] (node2){};
    \node [on chain=1, entry, fill=assumptions_asserted] (node3){\textcolor{assert}{$\bullet$}};
    \node [on chain=1, entry, fill=assumptions_asserted] (node4){\textcolor{assert}{$\bullet$}};
    \node [on chain=1, entry, draw=none] (empty2){};
\end{scope}
\draw[very thick,->, bend right=60, color=arrow_assumption, dashed] (node2.north) to (node1.north);
\draw[very thick,->, bend right=60, color=arrow_stream] (node3.north) to (node2.north);
\draw[very thick,->, bend left=60, color=arrow_stream] (node3.north) to (node4.north);
\draw[very thick,->, bend left=60, color=arrow_assumption, dashed] (node4.north) to (empty2.north);
\end{tikzpicture}
\caption{A stream execution of length $N+1$ with the corresponding template unfolding is depicted. 
The stream execution covers all cases where future out-of-accesses occur. 
Gray-colored and red dotted positions represent unfolded assumptions and assertions, respectively. 
Solid and dashed arrows indicate accesses by streams and annotations, respectively.}
\label{fig:endtrace}
\end{figure}
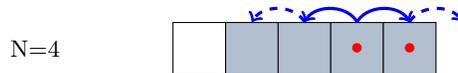

Last, we consider the case where only future out-of-bounds accesses occur.
Hence, the respective obligations need to incorporate default values of future out-of-bounds accesses.
As before, we can use the previously proven assertions as assumptions.
Figure~\ref{fig:endtrace} depicts a stream execution with two dotted positions, \ie positions where the assertion must be proven.
The position where arrows are given represents the case where only the assumption results in a future out-of-bounds access.
The last position of the stream execution represents the case in which both the assumption and the stream result in future out-of-bounds accesses.   
The presented unfolding is formalized as proof obligations in Definition~\ref{def_end}.

\begin{definition}[Proof Obligations for Future Out-of-bounds Accesses]\label{def_end}~\\
The proof obligations $\phi_{\mathit{End}}$ for future out-of-bounds accesses are defined as the template formula $\phi_{t}(\mathit{c\_asm}, ~\mathit{c\_asserted}, ~\mathit{c\_streams}, ~\mathit{c\_assert})$ with template parameters:\\
	\begin{tabular}{ll}
		 $~\bullet~ \mathit{c\_asm}$ & $:= w_p \leq i \leq N$,\\
		 $~\bullet~ \mathit{c\_asserted}$ & $:= 2 \cdot w_p \leq i < 3\cdot w_p$,\\
		 $~\bullet~ \mathit{c\_streams}$ & $:= 2 \cdot w_p \leq i \leq N$,\\
		 $~\bullet~ \mathit{c\_assert}$ & $:= 3\cdot w_p \leq i \leq N$
	\end{tabular}
	
\noindent where $N = 3  \cdot w_p + w_f$.
\end{definition}

So far, we have defined proof obligations for certain positions in the stream execution with and without out-of-bounds accesses.
Together, the proof obligations constitute an inductive argument for the correctness of the assertions, see Proposition~\ref{def-efficient}.
Here, the base case is given by Definition~\ref{def_begin} and induction steps are given by Definitions~\ref{def_run} and~\ref{def_end}. 
The induction steps use the induction hypothesis, \ie valid assertions, due to the template parameter $c\_asserted$.

\begin{proposition}[Assertion Verification by \lola Unfolding]\label{def-efficient}~\\
The set of assertions is correct if the formula $\phi_{\mathit{Begin}} ~\wedge~ \phi_{\mathit{Run}} ~\wedge~ \phi_{\mathit{End}}$ is valid.
\end{proposition}

Proposition~\ref{def-efficient} proves the soundness of the verification approach.
Soundness refers to the ability of an analyzer to prove the absence of errors --- if a \lola specification is accepted, it is guaranteed that the assertions are not violated.
The converse does not hold, \ie the presented verification approach is not complete.
Completeness refers to the ability of an analyzer to prove the presence of errors --- if a \lola specification is rejected, the counter-example given should be a valid stream execution that results in an assertion violation.
The following \lola specification is rejected even though no assertion is violated:
\begin{lstlisting}[escapechar=@]
input a: Int32
assume <a1> a @$\le$@ 10
output sum := if sum[-1, 0] @$\le$@ 10 then 0 else sum[-1, 0] + a
assert < a1 > sum @$\le$@ 100
\end{lstlisting} 
Here, since the $\mathtt{if}$-condition in Line 3 evaluates to $true$ at the beginning of the stream execution,  $\mathtt{sum}$ is a constant stream with value zero.
Hence, the assertion in Line 4 is never violated.
The verification approach rejects this specification.
The reason for this is that $\mathtt{sum} \le 100$ is added as an \emph{asserted} condition in $\phi_{\mathit{Run}}$.
Therefore, the \smt solver can assign a value between $91$ and $100$ to the earliest $\mathtt{sum}$ variable of the unfolding, resulting in an assertion violation of the next $\mathtt{sum}$ variable.

\vspace{5mm}

\begin{figure}
\centering
\begin{tikzpicture}
\edef\sizetape{0.7cm}
\tikzstyle{entry}=[draw,minimum size=\sizetape]
\begin{scope}[start chain=1 going right,node distance=-0.15mm]
    \node [on chain=1] {N=6};
    \node [on chain=1, entry, draw=none] (empty0){};
    \node [on chain=1, entry, draw=none] (empty1){};
    \node [on chain=1, entry] (node0){};
    \node [on chain=1, entry, fill=assumptions_asserted] (node1){};
    \node [on chain=1, entry, fill=assumptions_asserted] (node2){};
    \node [on chain=1, entry, fill=assumptions_asserted] (node3){\textcolor{assert}{$\bullet$}};
    \node [on chain=1, entry, fill=assumptions_asserted] (node4){};
    \node [on chain=1, entry, fill=assumptions_asserted] (node5){};
    \node [on chain=1, entry] (node6){};
\end{scope}
\draw[very thick,->, bend right=60, color=arrow_assumption, dashed] (node2.north) to (node1.north);
\draw[very thick,->, bend right=60, color=arrow_stream] (node3.north) to (node2.north);
\draw[very thick,->, bend left=60, color=arrow_stream] (node3.north) to (node4.north);
\draw[very thick,->, bend left=60, color=arrow_assumption, dashed] (node4.north) to (node5.north);
\end{tikzpicture}
\caption{A stream execution of length $N+1$ with the corresponding template unfolding is depicted.
The stream execution considers the case where no out-of-bound access occurs. 
Gray-colored and red dotted positions represent unfolded assumptions and assertions, respectively. 
Solid and dashed arrows indicate accesses by streams and annotations, respectively.}
\label{fig:runtrace}
\end{figure}
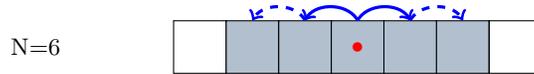

\newpage
\section{Application Experience in Avionics}\label{sec:experimental}In this section, we present details about the tool implementation and tool experiences on practical avionic specifications.

\paragraph{\textbf{Tool Implementation and Usage}}~\\
The tool is based on the open source \lola framework\footnote{https://rtlola.org/} written in Rust.
Specifically, it uses the \lola frontend to parse a given specification into an intermediate representation.
Based on this representation, the \smt formulas are created and evaluated with the Rust z3 crate\footnote{https://docs.rs/z3/0.9.0/z3/}.
At its current phase of the crate's development, a combined solver is implemented that internally uses either a non-incremental or an incremental solver.
There is no information on the implemented tactics available, but all our requests could be solved within seconds.
For functions that are not natively supported by the Rust Z3 solver, the output is arbitrarily chosen by the solver with respect to the range of the function.
The tool expects a \lola specification augmented by \emph{assumptions} and \emph{assertions}.
The verification is done automatically and produces a counter-example stream execution, if any exists.
The counter-example can then be used by the user to debug its specifications.
Two different kinds of users are targeted.
First, users that write the entire augmented specification.
Such a user could be a systems engineer who is developing a safety monitor and wants to ensure that it contains critical properties.
Second, users that augment an existing specification.
Here, one reason could be that an existing monitor shall be composed with other critical components and certain behavioral properties are expected.
Also, similar to software testing, the task of writing a specification and their respective assumptions and assertions could be separated between two users to ensure the independence of both.

\paragraph{\textbf{Practical Results}}~\\
To gain practical tool experience, previously written specifications based on interviews with engineers of the German Aerospace Center~\cite{schirmer2016} were extended by assumptions and assertions.
The previous specifications were tested using log-files and simulations -- the authors considered them correct.
We report several specification errors in Table \ref{tab:specification} that were detected by the presented verification extension.
In fact, the detected errors would have resulted in undetected failures. 
After the errors in the previous specifications were fixed, all assertions were proven correct.
Note that the errors could have been found due to manual reviews.
However, such reviews are tedious and error-prone, especially when temporal behaviors are involved.
The detected errors in Table \ref{tab:specification} can be grouped into three classes: \emph{Classical Bugs}, \emph{Operator Errors}, and \emph{Wrong Interpretations}.
Classical bugs are errors that occur when implementing an algorithm.
Operator errors are \lola specific errors, \eg temporal accesses.
Last, wrong interpretations refer to gaps between the specification and the user's design intend, \eg violated assertions due to incomplete specifications.
Next, we give one representative example for each group.
We reduced the specification to the representative fragment.

\begin{table}[t]
\centering
\begin{tabular}{|l|c|c|c|l|}
\hline
\textbf{Specification} & \textbf{\#o} & \textbf{\#a} & \textbf{\#g} & \textbf{Detected errors} \\
\hline
$\mathit{gps\_vel\_output}$ & 14 & 6 & 6 & \quad -- \\
\hline
$\mathit{gps\_pos\_output}$ & 19 & 3 & 10 & \quad -- \\
\hline
$\mathit{imu\_output}$ & 18 & 6 & 6 & Wrong default value \\
 &  &  &  & Division by zero \\
\hline
$\mathit{nav\_output}$ & 25 & 3 & 5 & Missing abs() \\
\hline
$\mathit{tagging}$ & 6 & 2 & 2 & \quad -- \\
\hline
$\mathit{ctrl\_output}$ & 25 & 7 & 8 & Wrong threshold comparisons \\
\hline
$\mathit{mm\_output\_1}$ & 4 & 1 & 2 & \quad -- \\
\hline
$\mathit{mm\_output\_2}$ & 17 & 6 & 9 & Missing if condition\\
 &  &  &  & Wrong default value \\
\hline
$\mathit{contingency\_output}$ & 4 & 8 & 1 & Observation: both contingencies could \\
					 &  &    &   & be true in case of voting, \ie both at 50\% \\
\hline
$health\_output$ & 1 & 5 & 1 & \quad -- \\
\hline
\end{tabular} 
\vspace{3mm}
\caption{Detected errors by the verification extension, where \#o, \#a, and \#g represent the number of outputs, assumptions, and assertions given in the specification, respectively.}
\label{tab:specification}
\end{table}

\begin{example}[Classical Bug]~\\
The \lola specification in Listing \ref{lst:expr_ctrl_output} monitors the fuel level. 
A monitor shall notify the operator when one of the three different fuel levels are reached: half (Line 8), warning (Line 9), and danger (Line 10).
The fuel level is computed as a percentage in Line 7.
It uses the fuel level at the beginning of the flight (Line 6) as a reference for its computation.
Given the documentation of the fuel sensor, it is known that \texttt{fuel} values are within $\mathbb{R}^{+}$ and decreasing.
This is formalized in Line 4 as an assumption.
As an invariant, we asserted that the starting fuel is greater or equal to \texttt{fuel} (Line 15).
Further, in Lines 16 to 18, we stated that once a level is reached it should remain at this level.
During our experiment, the assertion led to a counter-example that pointed to the previously used and erroneous fuel level computation: 
\begin{lstlisting}[escapechar=@, numbers=none]
output fuel_level := (start_fuel - fuel) / start_fuel
\end{lstlisting}
In short, the output computed the consumed fuel and not the remaining fuel.
The computation could be easily fixed by converting consumed fuel into remaining fuel, see Line 7.
Therefore, Listing \ref{lst:expr_ctrl_output} satisfies its assertion.
Note, that offset accesses were used to assert the temporal behavior of the fuel level output stream.
Further, \texttt{trigger\_once} is an abbreviation which states that only the first raising edge is reported to the user.

\begin{figure}[t]
\lstset
{ 
    numbers=right,
    stepnumber=1,
}
\begin{lstlisting}[escapechar=@, caption={The fixed version of the \lola ctrl\_output specification that monitors the fuel level. Three level of engagement are depicted: half, warning, and danger.}, label={lst:expr_ctrl_output}]
// Inputs 
input fuel: Float64
// Assumptions
assume<a5> fuel  > 0.0 and fuel  <  fuel[-1, fuel + 0.1]  
// Outputs
output start_fuel := start_fuel[-1, fuel] 
output fuel_level := @{\color{red}1.0}@ - (start_fuel - fuel) / start_fuel
output fuel_half    := fuel_level < 0.50
output fuel_warning := fuel_level < 0.25
output fuel_danger  := fuel_level < 0.10
trigger_once fuel_half    "INFO: Fuel level is half reduced"
trigger_once fuel_warning "WARNING: Fuel level is below 25%"
trigger_once fuel_danger  "DANGER: Fuel level is below 10%"
// Assertions
assert<a5>     start_fuel >= fuel 
           and (fuel_half[-1, false] -> fuel_half) 
           and (fuel_warning[-1, false] -> fuel_warning)
           and (fuel_danger[-1, false] -> fuel_danger)
\end{lstlisting} 
\end{figure}
\end{example}

\newpage
\begin{example}[Operator Error]~\\
An important monitoring property is to detect frozen values as these indicate a deteriorated sensor.
Such a specification is depicted in Listing \ref{lst:expr_imu_output}.
Here, as an input, the acceleration in $x-$direction is given.
The frozen value check is computed from Line 6 to Line 10.
It compares previous values using \lola's offset operator.
To check this computation, we added the sanity check that asserts that no frozen value shall be detected (Line 13) when small changes in the input are present (Line 4).
In the previous version, the frozen values were computed using the abbreviated offset operator:
\begin{lstlisting}[escapechar=@, numbers=none]
output frozen_ax := ax[-5..0, 0.0, =]
\end{lstlisting}
This resulted in a counter-example that pointed to wrong default values.
Although the abbreviated version is easier to read and reduces the size of the specification, it is unfortunately not suitable for this kind of property.
The tool detected the unlikely situation that the first value of \texttt{ax} is $0.0$ which would have resulted in evaluating \texttt{frozen\_ax} to true.
Although unlikely, this should be avoided as contingencies activated in such situations depend on correct results and otherwise could harm people on the ground.
By unfolding the operator and adding a different default value to one of the past accesses, the error was resolved (Line 6).
Listing \ref{lst:expr_imu_output} shows the fixed version which satisfies its assertion.

\begin{figure}[]
\lstset
{ 
    numbers=right,
    stepnumber=1,
}
\begin{lstlisting}[escapechar=@, caption={The \lola imu\_output specification that monitors frozen acceleration values.}, label={lst:expr_imu_output}]
// Inputs
input ax: Float32
// Assumptions
assume <a1> ax != ax[-1, ax + @{$\varepsilon$}@]  
// Outputs
output frozen_ax :=    ax[-5, @{\color{red} 0.1}@] = ax[-4, 0.0]            
                  and ax[-4, 0.0] = ax[-3, 0.0]   
                  and ax[-3, 0.0] = ax[-2, 0.0]  
                  and ax[-2, 0.0] = ax[-1, 0.0]  
                  and ax[-1, 0.0] = ax          
trigger frozen_ax "WARNING: x-acceleration is frozen!"
// Assertions
assert <a1> !frozen_ax
\end{lstlisting}
\end{figure}
\end{example}

\newpage
\begin{example}[Wrong Interpretation]~\\
In Listing \ref{lst:expr_contingency_output}, two visual sensor readings are received (Lines 2-3).
Both, readings argue over the same observations where \texttt{avgDist} represents the average distance to the measured obstacle, \texttt{actual} is the number of measurements, and \texttt{static} is the number of unchanged measurements.
A simple rating function is introduced (Lines 5-8) that estimates the corresponding rating -- the higher the better.
Using these ratings, the trust in each of the sensors is computed probabilistically (Lines 9-10).
When considering the integration of such a monitor as an ASTM switch condition that decides which sensor value should be forwarded, the specification should be revised.
This is the case because the assertion in Line 14 produces a counter-example which indicates that both trust triggers (Lines 11-12) can be activated at the same time.
A common solution for this problem is to introduce a priority between the sensors.
{
\lstset
{ 
    numbers=right,
    stepnumber=1,
}
\begin{lstlisting}[escapechar=@, caption={The \lola contingency\_output specification that uses an heuristic to decide which sensor is more trustworthy.}, label={lst:expr_contingency_output}]
// Inputs
input avgDist_laser, actual_laser, static_laser: Float64
input avgDist_optical, actual_optical, static_optical: Float64
// Outputs
output rating_laser := 
    0.2 * static_laser + 0.4 * actual_laser + 0.4 * avgDist_laser
output rating_optical := 
    0.2 * static_optical + 0.4 * actual_optical + 0.4 * avgDist_optical
output trust_laser := rating_laser / ( rating_laser + rating_optical)
output trust_optical := 1.0 - trust_laser
trigger trust_laser >= 0.5
trigger trust_optical >= 0.5
// Assertions
assert <a1> trust_laser != trust_optical
\end{lstlisting}
}
\end{example}

The examples show how the presented \lola verification extension can be used to find errors in specifications.
We also noticed that the annotations can serve as documentation.
System assumptions are often implicitly known during development and are finally documented in natural language in separate files.
Having these assumptions explicitly stated within the monitor specification potentially reduces future mistakes when reusing the specification, \eg when composing with other monitor specifications.
Listing~\ref{lst:simpledoc} depicts such an example specification.
Here, the monitor interfaces are clearly defined by the domain of input $a$ (Line 5) and output $o$ (Line 13).
Also, $\mathit{reset}$ is assumed to be valid at least once per second (Line 5).
Further, no deeper understanding of the internal computations (Lines 7-10) is required in order to safely compose this specification with others.

{
\lstset
{ 
    numbers=right,
    stepnumber=1,
}
\begin{lstlisting}[escapechar=@, caption={\lola specification annotations describe interface properties.}, label={lst:simpledoc}]
// Inputs with frequency 5Hz
input a: Float64
input reset: Bool
// Assumptions
assume <a1> 0.0 $\le$ a $\le$ 1.0 and reset[-4..0, false, $\vee$]
// Outputs
output o_1 := ...
...
output o_n :=  ...
output o := o_1 + ... + o_n
trigger o $\ge$ 0.5 "Warning: Output o exceeds threshold!"
// Assertions
assert <a1> 0.0 $\le$ o $\le$ 1.0
\end{lstlisting}
}

\section{Conclusion}\label{sec:conclusion}
As both the relevance and the complexity of cyber-physical systems continues to grow, runtime monitoring is an essential ingredient of safety-critical systems.
When monitors are derived from specifications it is crucial that the specifications are correct.
In this paper, we have presented a verification approach for the stream-based monitoring language \lola.
With this approach, the developer can formally prove guarantees on the streams computed by the monitor, and hence ensure that the monitor does not cause dangerous situations.
The verification extension is motivated by upcoming aviation regulations and standards as well as by practical feedback of engineers.

The extension has been applied to previously written \lola specifications that were obtained based on interviews with aviation experts. In this process, we discovered and fixed several serious specification errors.

In the future, we plan to develop automatic invariant generation for \lola specifications. Another interesting direction for future work is to exploit the results of the analysis for the optimization of the specification and the resulting monitoring code. 
Finally, we plan to extend the verification approach to \mbox{\textsc{RTLola}}, the real-time extension of \lola.



\section*{Acknowledgement}
This work was partially supported by the German Research Foundation (DFG) as part of the Collaborative Research Center Foundations of Perspicuous Software Systems (TRR 248, 389792660), by the European Research Council (ERC) Grant OSARES (No.~683300), and by the Aviation Research Programm LuFo of the German Federal Ministry for Economic Afairs and Energy as part of ``Volocopter Sicherheits-Technologie zur robusten eVTOL Flugzustandsabsicherung durch formales Monitoring''(No.~20Q1963C).
The final authenticated publication is available online at \url{https://doi.org/10.1007/978-3-030-88494-9_4}.




\bibliographystyle{splncs04}
\bibliography{bibliography}

\newpage
\appendix
\section{Lola Specifications -- Experience Report}\label{appendix:specs}

\subsection{$Specification: gps\_vel\_output$}\label{appendix:gps_vel_output}
\lstset
{ 
    numbers=right,
    stepnumber=1,
    basicstyle=\small,
}

\begin{lstlisting}[label={lst:gps_vel_output}]
input sol_age: Float32
input hor_spd: Float32 
input trk_gnd: Float32
input vert_spd: Float32
input time_s: UInt64
input time_us: UInt64
// Assumptions
assume <a1>     time - time.offset(by: -1).defaults(to: time - 0.1) > 0.0   
  and time - time.offset(by: -1).defaults(to: time - 0.1) <= 0.1 
  and trace_pos >= 0

assume <a2>     time - time.offset(by: -1).defaults(to: time - 0.1) > 0.0   
  and time - time.offset(by: -1).defaults(to: time - 0.1) <= 0.1 
  and trace_pos >= 0
// Frequency computations
output time := cast(time_s) + cast(time_us) / 1000000.0
output start_time := if time.offset(by: -1).defaults(to: -1.0) = -1.0 then time else start_time.offset(by: -1).defaults(to: -1.0) 
output flight_time := time - start_time
output trace_pos @ sol_age or hor_spd or  trk_gnd or vert_spd or time_s or time_us := trace_pos.offset(by: -1).defaults(to: -1) + 1
output frequency := 
     1.0 / ( time - time.offset(by: -1).defaults(to: time - 0.0001) )  
output freq_sum := 
     freq_sum.offset(by: -1).defaults(to: 0.0)  + frequency
output freq_avg := freq_sum / cast(trace_pos+1)
output freq_max := if frequency > freq_max.offset(by: -1).defaults(to: frequency)  then frequency else freq_max.offset(by: -1).defaults(to: frequency)
output freq_min := if frequency < freq_min.offset(by: -1).defaults(to: frequency)  then frequency else freq_min.offset(by: -1).defaults(to: frequency)
// Speed computations
output hor_spd_max := if hor_spd > hor_spd_max.offset(by: -1).defaults(to: 0.0)  then hor_spd else hor_spd_max.offset(by: -1).defaults(to: 0.0)
output vert_spd_max := if vert_spd > vert_spd_max.offset(by: -1).defaults(to: 0.0)  then vert_spd else vert_spd_max.offset(by: -1).defaults(to: 0.0)
// Solution age and track over ground (motion direction wrt. north)
trigger sol_age <= 0.5  "Sol age should remain zero!"
output trk_gnd_in_bound := if trk_gnd >= 0.0 and trk_gnd <= 360.0 then trk_gnd_in_bound.offset(by: -1).defaults(to: true)  else  false 
output trk_gnd_max := if trk_gnd > trk_gnd_min.offset(by: -1).defaults(to: 0.0)  then trk_gnd else trk_gnd_min.offset(by: -1).defaults(to: 0.0)
output trk_gnd_min := if trk_gnd < trk_gnd_max.offset(by: -1).defaults(to: 0.0)  then trk_gnd else trk_gnd_max.offset(by: -1).defaults(to: 0.0)
// Assertions
assert <a1> time.offset(by: -1).defaults(to: -1.0) < time  
  and start_time == start_time.offset(by: -1).defaults(to: start_time) 
  and flight_time >= flight_time.offset(by: -1).defaults(to: 0.0)  
assert <a2>     frequency >= 10.0
  and freq_sum >= freq_sum.offset(by: -1).defaults(to: 0.0) + 10.0      
assert <a3> trk_gnd_in_bound.offset(by: -1).defaults(to: true) 
  or !trk_gnd_in_bound 
\end{lstlisting}

\subsection{$Specification: gps\_pos\_output$}\label{appendix:gps_pos_output}
\lstset
{ 
    numbers=right,
    stepnumber=1,
    basicstyle=\small,
}

\begin{lstlisting}[label={lst:gps_pos_output}]
import math
input lat: Float32
input lon: Float32
input hgt: Float32
input nObjs: UInt64
input nGPSL1: UInt64
input time_s: UInt64
input time_us: UInt64
// Assumptions
assume <a1>     time - time.offset(by: -1).defaults(to: time - 0.1) > 0.0   
  and time - time.offset(by: -1).defaults(to: time - 0.1) <= 0.1 
  and trace_pos >= 0
// Frequency computations
output time: Float32 := cast(time_s) + cast(time_us) / 1000000.0
output start_time := if time.offset(by: -1).defaults(to: -1.0) == -1.0 then time else start_time.offset(by: -1).defaults(to: -1.0) 
output flight_time := time - start_time
output trace_pos @ lat or lon or hgt or nObjs or nGPSL1 or time_s or time_us := trace_pos.offset(by: -1).defaults(to: -1) + 1 
output frequency := 1.0 / ( time - time.offset(by: -1).defaults(to: time - 0.0001) ) 
output freq_sum := freq_sum.offset(by: -1).defaults(to: 0.0)  + frequency
output freq_avg := freq_sum / cast(trace_pos+1)
output freq_max := if frequency > freq_max.offset(by: -1).defaults(to: 0.0)  then frequency else freq_max.offset(by: -1).defaults(to: 0.0)
output freq_min := if frequency < freq_min.offset(by: -1).defaults(to: 0.0)  then frequency else freq_min.offset(by: -1).defaults(to: 0.0)
// Statistics
output lat_max := if lat > lat_max.offset(by: -1).defaults(to: lat)  then lat else lat_max.offset(by: -1).defaults(to: lat)
output lat_min := if lat < lat_min.offset(by: -1).defaults(to: lat)  then lat else lat_min.offset(by: -1).defaults(to: lat)
output lon_max := if lon > lon_max.offset(by: -1).defaults(to: lon)  then lon else lon_max.offset(by: -1).defaults(to: lon)
output lon_min := if lon < lon_min.offset(by: -1).defaults(to: lon)  then lon else lon_min.offset(by: -1).defaults(to: lon)
output lat_in_bound := max( abs(lat_max), abs(lat_min) ) <= 90.0
output lon_in_bound := max( abs(lon_max), abs(lon_min) ) <= 180.0
trigger !lat_in_bound "Irregular latitude value!"
trigger !lon_in_bound "Irregular longitude value!"
output begin := false
output start_height := if begin.offset(by: -1).defaults(to: true) then hgt else start_height.offset(by: -1).defaults(to: 0.0) 
output hgt_inc_max := max( hgt_inc_max.offset(by: -1).defaults(to: 0.0),  hgt - start_height )
output hgt_dec_max := min( hgt_dec_max.offset(by: -1).defaults(to: 0.0) , hgt - start_height )
trigger hgt_inc_max > 100.0 "Never increase height by more than 100m!"
trigger hgt_dec_max < -100.0 "Never decrease height by more than 100m"
// Assertions
assert <a1>     time.offset(by: -1).defaults(to: -1.0) < time 
  and start_time == start_time.offset(by: -1).defaults(to: start_time) 
  and flight_time >= flight_time.offset(by: -1).defaults(to: 0.0)  
assert <a2>     hgt_inc_max >= 0.0 and hgt_dec_max <= 0.0
  and hgt_inc_max >= hgt_inc_max.offset(by: -1).defaults(to: 0.0) 
  and hgt_dec_max <= hgt_inc_max.offset(by: -1).defaults(to: 0.0) 
  and start_height = start_height.offset(by: -1).defaults(to: start_height)
  and (lat_in_bound.offset(by: -1).defaults(to: true) or !lat_in_bound)   
  and (lon_in_bound.offset(by: -1).defaults(to: true) or !lon_in_bound)   
\end{lstlisting}

\subsection{$Specification: imu\_output$}\label{appendix:imu_output}
\lstset
{ 
    numbers=right,
    stepnumber=1,
    basicstyle=\small,
}

\begin{lstlisting}[label={lst:imu_output}]
import math
input ax: Float32
input ay: Float32
input az: Float32
input time_s: UInt64
input time_us: UInt64
input counter: Int64
// Assumptions
assume <a1>     time - time.offset(by: -1).defaults(to: time - 0.1) > 0.0   
  and time - time.offset(by: -1).defaults(to: time - 0.1) <= 0.1 
  and trace_pos >= 0
assume <a2>     ax != ax.offset(by: -1).defaults(to: ax + 0.1)    
  and ay != ay.offset(by: -1).defaults(to: ay + 0.1)
  and az != az.offset(by: -1).defaults(to: az + 0.1) 
// Frequency computations
output time := cast(time_s) + cast(time_us) / 1000000.0
output start_time := if time.offset(by: -1).defaults(to: -1.0) == -1.0 then time else start_time.offset(by: -1).defaults(to: -1.0) 
output flight_time := time - start_time
output trace_pos @ ax or ay or az or time_s or time_us or counter := trace_pos.offset(by: -1).defaults(to: -1) + 1
output frequency := 1.0 / ( time - time.offset(by: -1).defaults(to: time - 0.0001) )  
output freq_sum := freq_sum.offset(by: -1).defaults(to: 0.0)  + frequency
output freq_avg := freq_sum / cast(trace_pos+1)
// Statistics
output deviation := abs( frequency - 100.0)
output exceeds_worst := deviation > worst_dev.offset(by: -1).defaults(to: 0.0) 
output worst_dev_pos := if exceeds_worst then trace_pos else worst_dev_pos.offset(by: -1).defaults(to: 0) 
output worst_dev := if exceeds_worst then deviation else worst_dev.offset(by: -1).defaults(to: 0.0) 
output ax_max := max(abs(ax),ax_max.offset(by:-1).defaults(to:0.0))
output ay_max := max(abs(ay),ay_max.offset(by:-1).defaults(to:0.0))
output az_max := max(abs(az),az_max.offset(by:-1).defaults(to:0.0))
trigger ax > 15.0 or ay > 15.0 or az > 15.0
output frozen_ax := ax.offset(by:-1).defaults(to:0.0) = ax
  and ax.offset(by:-2).defaults(to:0.0)=ax.offset(by:-1).defaults(to:0.0)  
  and ax.offset(by:-3).defaults(to:0.0)=ax.offset(by:-2).defaults(to:0.0)   
  and ax.offset(by:-4).defaults(to:0.0)=ax.offset(by:-3).defaults(to:0.0)  
  and ax.offset(by:-5).defaults(to:0.1)=ax.offset(by:-4).defaults(to:0.0)
output frozen_ay := ay.offset(by:-1).defaults(to:0.0) = ay
  and ay.offset(by:-2).defaults(to:0.0)=ay.offset(by:-1).defaults(to:0.0)  
  and ay.offset(by:-3).defaults(to:0.0)=ay.offset(by:-2).defaults(to:0.0)   
  and ay.offset(by:-4).defaults(to:0.0)=ay.offset(by:-3).defaults(to:0.0)  
  and ay.offset(by:-5).defaults(to:0.1)=ay.offset(by:-4).defaults(to:0.0)
output frozen_az :=  az.offset(by:-1).defaults(to:0.0) = az
  and az.offset(by:-2).defaults(to:0.0)=az.offset(by:-1).defaults(to:0.0)  
  and az.offset(by:-3).defaults(to:0.0)=az.offset(by:-2).defaults(to:0.0)   
  and az.offset(by:-4).defaults(to:0.0)=az.offset(by:-3).defaults(to:0.0)  
  and az.offset(by:-5).defaults(to:0.1)=az.offset(by:-4).defaults(to:0.0)
trigger frozen_ax or frozen_ay or frozen_az
output check_counter := if trace_pos = 0 then false else (counter != (counter.offset(by: -1).defaults(to: -1) + 1) % 100) 
trigger check_counter  "A counter value was ignored."
// Assertions
assert <a1>     time.offset(by: -1).defaults(to: -1.0) < time  
  and start_time == start_time.offset(by: -1).defaults(to: start_time) 
  and flight_time >= flight_time.offset(by: -1).defaults(to: 0.0)  
assert <a2> !frozen_ax and !frozen_ay and !frozen_az
\end{lstlisting}

\subsection{$Specification: nav\_output$}\label{appendix:nav_output}
\lstset
{ 
    numbers=right,
    stepnumber=1,
    basicstyle=\small,
}

\begin{lstlisting}[label={lst:nav_output}]
import math
input lat: Float32
input lon: Float32
input ug: Float32
input vg: Float32
input wg: Float32
input time_s: UInt64
input time_us: UInt64
// Assertion
assume <a1>    trace_pos >= 0   
  and time - time.offset(by: -1).defaults(to: time - 0.1) <= 0.1 
  and time - time.offset(by: -1).defaults(to: time - 0.1) > 0.0 
// Frequency Computation
output time := cast(time_s) + cast(time_us) / 1000000.0
output start_time := if time.offset(by: -1).defaults(to: -1.0) == -1.0 then time else start_time.offset(by: -1).defaults(to: -1.0) 
output flight_time := time - start_time
output trace_pos @lat or lon or ug or vg or wg or time_s or time_us := trace_pos.offset(by: -1).defaults(to: -1) + 1 
output frequency :=  1.0 / ( time - time.offset(by: -1).defaults(to: time - 0.0001) )  
output freq_sum := freq_sum.offset(by: -1).defaults(to: 0.0)  + frequency
output freq_avg := freq_sum / cast(trace_pos+1)
output freq_max := if frequency > freq_max.offset(by: -1).defaults(to: frequency)  then frequency else freq_max.offset(by: -1).defaults(to: frequency)
output freq_min := if frequency < freq_min.offset(by: -1).defaults(to: frequency)  then frequency else freq_min.offset(by: -1).defaults(to: frequency)
// Statistics
output velocity := sqrt( ug*ug + vg*vg + wg*wg)
output lon1_rad := lon.offset(by: -1).defaults(to: 0.0)  * 3.1415926535 / 180.0
output lon2_rad := lon * 3.1415926535 / 180.0
output lat1_rad := lat.offset(by: -1).defaults(to: 0.0)  * 3.1415926535 / 180.0
output lat2_rad := lat * 3.1415926535 / 180.0
output dlon := lon2_rad - lon1_rad
output dlat := lat2_rad - lat1_rad
output a := (sin(dlat/2.0))*(sin(dlat/2.0)) + cos(lat1_rad) * cos(lat2_rad) * (sin(dlon/2.0))*(sin(dlon/2.0))
output x_atan2 := sqrt(a)
output y_atan2 := sqrt(1.0-a)
output c := 2.0 * if x_atan2 > 0.0                     then arctan(y_atan2/x_atan2)
  else if x_atan2 < 0.0 and y_atan2 >= 0.0  
    then arctan(y_atan2/x_atan2) + 3.1415926535
  else if x_atan2 < 0.0 and y_atan2 <  0.0  
    then arctan(y_atan2/x_atan2) - 3.1415926535
  else if x_atan2 = 0.0 and y_atan2 > 0.0   then  3.1415926535 / 2.0
  else if x_atan2 = 0.0 and y_atan2 < 0.0   then -3.1415926535 / 2.0
  else 0.0
output gps_distance := 6373000.0 * c
output passed_time := time - time.offset(by: -1).defaults(to: 0.0) 
output distance_max := velocity * passed_time
output dif_distance := abs( gps_distance - distance_max )  
output detected_jump :=if trace_pos=0 then false else dif_distance>1
trigger detected_jump "Jump!"
// Assertions
assert <a1>     time.offset(by: -1).defaults(to: -1.0) < time  
  and start_time == start_time.offset(by: -1).defaults(to: start_time) 
  and flight_time >= flight_time.offset(by: -1).defaults(to: 0.0) 
assert <a2>   (!detected_jump or gps_distance > distance_max) 
  or (!detected_jump or distance_max > gps_distance)
\end{lstlisting}

\subsection{$Specification: tagging$}\label{appendix:tagging}
\lstset
{ 
    numbers=right,
    stepnumber=1,
    basicstyle=\small,
}

\begin{lstlisting}[label={lst:tagging}]
import math 
input time_s: UInt64
input time_us: UInt64
input vel: Float64
// Assumptions
assume<a1>      (time_s = time_s.offset(by: -1).defaults(to: 0) 
  and time_us > time_us.offset(by: -1).defaults(to: 0))
  and (   time_s > time_s.offset(by: -1).defaults(to: 0) 
       or time_us > time_us.offset(by: -1).defaults(to: 0))
// Exemplary State Statistics
output time := cast(time_s) + cast(time_us) / 1000000.0
output correct_vel := abs( vel ) < 0.3
output cur_state := if correct_vel then
  if cur_state.offset(by: -1).defaults(to: 0) = 0 then 1 else 2 else 0                  
output start_interval := cur_state = 2 
output interval_start := if start_interval then interval_start.offset(by: -1).defaults(to: 0.0) else time
trigger start_interval "Interval started!"
output end_interval := cur_state.offset(by: -1).defaults(to: 0)  > 0 and !correct_vel and time_since_start >  5.0
trigger end_interval "Interval ended!"
output time_since_start := time - interval_start.offset(by: -1).defaults(to: 0.0) 
// Assertions
assert <a1>     !(start_interval and end_interval) 
  and time_since_start > 0.0
\end{lstlisting}

\subsection{$Specification: ctrl\_output$}\label{appendix:ctrl_output}
\lstset
{ 
    numbers=right,
    stepnumber=1,
    basicstyle=\small,
}

\begin{lstlisting}[label={lst:ctrl_output}]
import math
input time_s: UInt64
input time_us: UInt64
input vel_x: Float64
input vel_y: Float64
input vel_z: Float64
input fuel: Float64
input power: Float64
input vel_r_x: Float64
input vel_r_y: Float64
input vel_r_z: Float64
// Assumptions
assume <a1>     trace_pos >= 0
  and time - time.offset(by: -1).defaults(to: time - 0.1) <= 0.1 
  and time - time.offset(by: -1).defaults(to: time - 0.1) > 0.0   
assume<a2>      power > 0.0 
  and power <= power.offset(by: -1).defaults(to: power)  
  and fuel  > 0.0 and fuel  <  fuel.offset(by: -1).defaults(to: fuel + 0.1) 
  and (time_s = time_s.offset(by: -1).defaults(to: 0) 
  and time_us > time_us.offset(by: -1).defaults(to: 0))
  and (time_s > time_s.offset(by: -1).defaults(to: 0) 
       or time_us > time_us.offset(by: -1).defaults(to: 0))
// Frequency computations
output time := cast(time_s) + cast(time_us) / 1000000.0
output start_time := if time.offset(by: -1).defaults(to: -1.0) == -1.0 then time else start_time.offset(by: -1).defaults(to: -1.0) 
output flight_time := time - start_time
output trace_pos @ time_s or time_us or vel_x or vel_y or vel_z or fuel or power or vel_r_x or vel_r_y or vel_r_z := trace_pos.offset(by:-1).defaults(to:-1) + 1
output frequency := 1.0 / ( time - time.offset(by: -1).defaults(to: time - 0.0001) )  // major improvement
output freq_sum := freq_sum.offset(by: -1).defaults(to: 0.0)  + frequency
output freq_avg := freq_sum / cast(trace_pos+1)
output freq_max := if frequency > freq_max.offset(by: -1).defaults(to: frequency)  then frequency else freq_max.offset(by: -1).defaults(to: frequency)
output freq_min := if frequency < freq_min.offset(by: -1).defaults(to: frequency)  then frequency else freq_min.offset(by: -1).defaults(to: frequency)
// Exemplary phase detection
output velocity := sqrt( vel_x*vel_x + vel_y*vel_y + vel_z*vel_z )
output velocity_max := if reset_max.offset(by: -1).defaults(to: false) then velocity else max( velocity, velocity_max.offset(by: -1).defaults(to: 0.0) ) 
output velocity_min := if reset_max.offset(by: -1).defaults(to: false) then velocity else min( velocity, velocity_min.offset(by: -1).defaults(to: 0.0) ) 
output dif_max := abs(velocity_max - velocity_min)
output reset_max := dif_max > 1.0
output reset_time := if reset_max or trace_pos = 0 then time else reset_time.offset(by: -1).defaults(to: 0.0)
output unchanged := if reset_max.offset(by: -1).defaults(to: false) then 0 else unchanged.offset(by: -1).defaults(to: 0) + 1 
trigger unchanged = 150 "Phase detected!"
// Statistics
output vel_dev := abs(vel_r_x-vel_x) + abs(vel_r_y-vel_y) + abs(vel_r_z-vel_z)
output dev_sum := vel_dev + dev_sum.offset(by: -1).defaults(to: 0.0) 
output vel_av := dev_sum / cast((trace_pos+1)*3)
output worst_dev_pos := if worst_dev.offset(by: -1).defaults(to: vel_dev - 1.0) < vel_dev then trace_pos else worst_dev_pos.offset(by: -1).defaults(to: 0) 
output worst_dev := if worst_dev.offset(by: -1).defaults(to: vel_dev - 1.0) < vel_dev then vel_dev else worst_dev.offset(by: -1).defaults(to: 0.0) 
output start_fuel := start_fuel.offset(by: -1).defaults(to: fuel) 
output fuel_level := 1 - ( start_fuel - fuel ) / start_fuel
output fuel_half    := fuel_level < 0.50
output fuel_warning := fuel_level < 0.25
output fuel_danger  := fuel_level < 0.10
output start_power := start_power.offset(by: -1).defaults(to: power) 
output power_p_consumed := ( (start_power - power) / (start_power) )
trigger_once fuel_half    "INFO: Fuel level is half reduced"
trigger_once fuel_warning "WARNING: Fuel level is below 25%"
trigger_once fuel_danger  "DANGER: Fuel level is below 10%"
trigger_once power_p_consumed > 0.50  "Power below half capacity"
trigger_once power_p_consumed > 0.75  "Power below quarter capacity"
trigger_once power_p_consumed > 0.90  "Urgent: Recharge Power!"
// Assertions
assert <a1>     time.offset(by: -1).defaults(to: -1.0) < time  
  and start_time == start_time.offset(by: -1).defaults(to: start_time) 
  and flight_time >= flight_time.offset(by: -1).defaults(to: 0.0)  
assert<a2>      reset_time >= 0.0 
  and start_fuel >= fuel and start_power >= power
  and (!fuel_half.offset(by: -1).defaults(to: false) or fuel_half) 
  and (!fuel_warning.offset(by: -1).defaults(to: false)or fuel_warning)
  and (!fuel_danger.offset(by: -1).defaults(to: false) or fuel_danger) 
  and power_p_consumed >= power_p_consumed.offset(by: -1).defaults(to: power_p_consumed) 
\end{lstlisting}

\subsection{$Specification: mm\_output\_1$}\label{appendix:mm_output_1}
\lstset
{ 
    numbers=right,
    stepnumber=1,
    basicstyle=\small,
}

\begin{lstlisting}[label={lst:mm_output_1}]
import math
input stateID_SC: UInt64
// Assumptions
assume<a1>  trace_pos >= 0
// Exemplary state transition analysis
output trace_pos @ stateID_SC := trace_pos.offset(by: -1).defaults(to: -1) + 1
output change_state := if trace_pos = 0 then false 
  else stateID_SC != stateID_SC.offset(by: -1).defaults(to: 0) 
output transitions := if stateID_SC.offset(by:-1).defaults(to: 0) = 0 then stateID_SC == 1 
  else if stateID_SC.offset(by: -1).defaults(to: 0) == 1 then stateID_SC == 1 or stateID_SC == 2
  else if stateID_SC.offset(by: -1).defaults(to: 0) == 2 then stateID_SC == 1 or stateID_SC == 3 
  else if stateID_SC.offset(by: -1).defaults(to: 0) == 3 then stateID_SC == 3
  else false
output invalid_transitions := change_state and !transitions
trigger invalid_transitions "Invalid state transition"
// Assertions
assert <a1> invalid_transitions or
!( stateID_SC.offset(by: -1).defaults(to: 0) != 0 and stateID_SC = 0 )  
assert <a2> (stateID_SC == 1 or stateID_SC == 2 or stateID_SC == 3 )   
  or !( stateID_SC.offset(by: -2).defaults(to: 0) = 1 
   and  transitions.offset(by: -1).defaults(to: false) and transitions )       
\end{lstlisting}

\subsection{$Specification: mm\_output\_2$}\label{appendix:mm_output_2}
\lstset
{ 
    numbers=right,
    stepnumber=1,
    basicstyle=\small,
}

\begin{lstlisting}[label={lst:mm_output_2}]
import math
input time_s: UInt64
input time_us: UInt64
input stateID_SC: Int64
input OnGround: UInt64
// Assumptions
assume <a1>     trace_pos >= 0
  and time - time.offset(by: -1).defaults(to: time - 0.1) <= 0.1 
  and time - time.offset(by: -1).defaults(to: time - 0.1) > 0.0   
assume <a2>     (time_s = time_s.offset(by: -1).defaults(to: 0) 
  and time_us > time_us.offset(by: -1).defaults(to: 0))
  and (time_s > time_s.offset(by: -1).defaults(to: 0) 
       or time_us > time_us.offset(by: -1).defaults(to: 0))
// Frequency computations
output time := cast(time_s) + cast(time_us) / 1000000.0
output start_time := if time.offset(by: -1).defaults(to: -1.0) == -1.0 then time else start_time.offset(by: -1).defaults(to: -1.0) 
output flight_time := time - start_time
output trace_pos @ time_s or time_us or stateID_SC or OnGround := trace_pos.offset(by: -1).defaults(to: -1) + 1
output frequency := 1.0 / ( time - time.offset(by: -1).defaults(to: time - 0.0001) )  
output freq_sum := freq_sum.offset(by: -1).defaults(to: 0.0)  + frequency
output freq_avg := freq_sum / cast(trace_pos+1)
// Phase Statistics
output change_state := if trace_pos = 0 then false 
       else stateID_SC != stateID_SC.offset(by: -1).defaults(to: 0) 
trigger change_state
output entrance_time := if change_state then time 
         else entrance_time.offset(by: -1).defaults(to: time) 
output hover_end := change_state and stateID_SC.offset(by: -1).defaults(to: -1)  = 4
output hover_cur_time := if hover_end then 
time - entrance_time.offset(by: -1).defaults(to: 0.0)else  0.0
output hover_sum_time := hover_sum_time.offset(by: -1).defaults(to: 0.0) + hover_cur_time
output hover_num_times := hover_num_times.offset(by: -1).defaults(to: 0)  + if hover_end then 1 else  0
output hover_max_time := max ( hover_max_time.offset(by: -1).defaults(to: 0.0), hover_cur_time )
output hover_avg_time := if hover_num_times != 0 then hover_sum_time / cast(hover_num_times) else 0.0
output landing_info := if change_state and stateID_SC = 5 then 0.0 else time - entrance_time.offset(by: -1).defaults(to: time) 
output landing_error := stateID_SC = 5 and OnGround != 1 and landing_info > 20.0
// Assertions
assert <a1>     time.offset(by: -1).defaults(to: -1.0) < time 
  and start_time == start_time.offset(by: -1).defaults(to: start_time) 
  and flight_time >= flight_time.offset(by: -1).defaults(to: 0.0)  
assert <a2>     time >= entrance_time and start_time <= entrance_time
  and hover_cur_time >= 0.0 and hover_max_time <= flight_time
assert <a3>     !(landing_error and hover_end) 
  and (!landing_error or landing_info > 0.0)
\end{lstlisting}

\subsection{$Specification: contingency\_output$}\label{appendix:contingency_output}
\lstset
{ 
    numbers=right,
    stepnumber=1,
    basicstyle=\small,
}

\begin{lstlisting}[label={lst:contingency_output}]
input avgDist_laser: Float64 
input actual_laser: Float64
input static_laser: Float64
input avgDist_optical:Float64 
input actual_optical: Float64
input static_optical: Float64
// Assumptions
assume <a1>     avgDist_laser >= 0.0 and actual_laser >= 0.0    
  and static_laser >= 0.0 and avgDist_optical >= 0.0    
  and actual_optical >= 0.0 and static_optical >= 0.0
  and (avgDist_laser + actual_laser + static_laser > 0.0)
  and (avgDist_optical + actual_optical + static_optical > 0.0)
// Trust computations
output rating_laser := 0.2 * static_laser + 0.4 * actual_laser       
     + 0.4 * avgDist_laser
output rating_optical := 0.2 * static_optical + 0.4 * actual_optical + 0.4 * avgDist_optical
output trust_laser := rating_laser / ( rating_laser + rating_optical)
output trust_optical := 1.0 - trust_laser
trigger trust_laser >= 0.5 "Trust in laser"
trigger trust_optical > 0.5 "Trust in optical sensor"
// Assertions
assert <a1> trust_laser $\neq$ trust_optical     
\end{lstlisting}

\subsection{$Specification: health\_output$}\label{appendix:health_output}
\lstset
{ 
    numbers=right,
    stepnumber=1,
    basicstyle=\small,
}
\tiny
\begin{lstlisting}[label={lst:health_output}]
import math
// average distance to the measured ostacle (range of sight) using laser
input avgDist_laser: Float64    
// average distance to the measured ostacle (range of sight) using camera
input avgDist_optical: Float64  
input vel: Float64
// Assumption
assume <a1>     avgDist_laser <= 100.0 and avgDist_laser >=   0.0   // both in m
  and avgDist_optical <= 50.0  and avgDist_optical >=  0.0   // both in m
  and abs(vel) < 5.5  // in m/s
// Line of sight
output avgDst_dif := min( avgDist_laser, avgDist_optical ) - abs(vel)
trigger avgDst_dif < 5.0 "WARNING: Dynamic Velocity Limit reached"
trigger avgDst_dif < 2.0 "ERROR: Abort mission."
// Assertions
assert <a1>     avgDst_dif < 54.5 and avgDst_dif > -5.5   
\end{lstlisting}

\end{document}